\documentclass[10pt]{iopart}

\usepackage{iopams} 
\usepackage{graphicx}
\usepackage[caption=false]{subfig}
\usepackage[export]{adjustbox}
\usepackage{placeins}
\usepackage[squaren]{SIunits}
\usepackage{hyperref}
\usepackage{iopams}
\usepackage[dvipsnames]{xcolor}

\definecolor{correction}{rgb}{1.0, 0.03, 0.0}
\definecolor{discuss}{rgb}{0.0, 0.28, 0.67}

\begin{document}

\title[]{Monte Carlo simulation of ultrafast nonequilibrium spin and charge transport in iron}

\author{J. Briones, H. C. Schneider  and B. Rethfeld}

\address{ Department of Physics and OPTIMAS Research Center, TU Kaiserslautern, Erwin Schroedinger Str. 46, 67663 Kaiserslautern, Germany}
\ead{briones@physik.uni-kl.de}
\vspace{10pt}

\begin{abstract}
Spin transport and spin dynamics after femtosecond laser pulse irradiation of iron (Fe) are studied using a kinetic Monte Carlo model. This model simulates spin dependent dynamics by taking into account two interaction processes during nonequilibrium: Elastic electron - lattice scattering, where only the direction of the excited electrons changes neglecting the energy loss, and inelastic electron - electron interaction, where secondary electrons are generated. An analysis of the particle kinetics inside the material shows that a smaller elastic scattering time affects the spin dynamics by leading to a larger spatial spread of electrons in the material, whereas generation of secondary electrons affects the spin transport with a larger time of propagation of homogeneous spin polarization.

\end{abstract}

%
%
%
%
\ioptwocol

\section{Introduction}

Following the discovery of ultrafast demagnetization in metallic ferromagnets and its connection to hot electron spin transport, the interplay of optical excitation, magnetization dynamics and transport is still under active investigation. For the optical excitation of ferromagnets the electronic system is brought out of equilibrium, which will be later thermalized to the surrounding environment with many possible scattering mechanisms \cite{PhysRevLett.111.167204, PhysRevB.90.014417, PhysRevLett.85.3025, Krieger_2017, PhysRevB.79.140401}. 


The dynamics of spin transport were first studied by Battiato, et al. \cite{PhysRevLett.105.027203} identifying an intermediate regime of the spin transport, labeled as superdiffusive transport. The contribution to demagnetization dynamics was later supported by experiments \cite{Rudolf, PhysRevLett.117.147203}. Another recent study of this effect introduced a particle in cell simulation \cite{PhysRevB.98.224416} to solve the Boltzmann equation for spin dependent hot-electron transport.

In order to understand the influence of the different scattering interactions in spin transport we will analyze the spatio-temporal dynamics  by tracing spin and charge in dependence of depth by considering free electron states for energies above the Fermi energy. For this reason we propose for this type of system an application of the kinetics Monte Carlo technique. 

Monte Carlo simulations have been widely used in studies of hot carriers dynamics, such as radiation biology \cite{Nikjoo_2008, Medvedev_2009, HUTHMACHER2015242}, nuclear physics \cite{Morales2009} and particle transport \cite{Zimmerman1991AlgorithmsFM} among many others. In this paper we present a Monte Carlo approach and its capabilities in analyzing the influence of secondary electrons generation in spin dynamics and spin transport. 

The outline of the paper is as follows: 
we will first present a model of excitation process, thereby briefly introduce how
the kinetic Monte Carlo technique works. We explain how an electron is treated during laser excitation and discuss the role of the spin dependent density of states. The next section will focus on the possible scattering processes that are taken into account in our simulation. 
Finally we present the results in depth-dependence for different scattering times in ferromagnetic iron after ultrafast laser excitation with $6\;\mathrm{eV}$ photon energy, as well as some results showing
the influence of secondary electrons generation in the particle kinetics. 

\section{Excitation Process}
\label{section::excitation}

In this section we discuss the algorithm used for the simulations presented in this paper, as well as an analysis of the energy density of excited electrons when only photoexcitation is considered.

\subsection{Monte Carlo algorithm}\label{subsection::MC_algorythm}

The asymptotic Monte Carlo trajectory method \cite{MC_asymp} is a statistical technique that models binary collision interactions by random sampling a very large number of trajectories until a result converges. The algorithm used for random sampling of a variable $x$ is done using probability theory. In probability theory one integrates the probabilities of all possible events $p(x)$, where $x_\mathrm{min} \leq x \leq x_\mathrm{max}$, into a variable called cumulative distribution function (CDF) $F(x)$ \cite{MCparticle}. One can map the CDF onto the range of random variables $\mathrm{R}$, where $\mathrm{R} \in [0,1]$ and R is distributed uniformly:
\begin{equation}
\label{eq:sampling}
 \mathrm{R} = \frac{F(x) - F(x_{min})}{F(x_{\mathrm{max}}) - F(x_{\mathrm{min}})} = \frac{\int \displaylimits_{x_{\mathrm{min}}}^{x} p(x)\, \mathrm{d}x }{\int \displaylimits_{x_{\mathrm{min}}}^{x_{ \mathrm{ max}}} p(x)\, \mathrm{d}x }\;,
\end{equation}
The variable $x$ is then uniquely determined in dependence on $\mathrm{R}$. Now we use this general idea to random-sampling any required  variable for the different types of interactions.

Random sampling is also used to decide whether certain interaction or excitation processes happen. In these cases we make use of the discrete form of eq.~(\ref{eq:sampling}). Examples of random sampling are shown in Figs.~\ref{fig:DOS_CDF} and \ref{fig:Prob_depth} for the excitation of electrons. The treatment of interactions in this simulation is discussed in 
section \ref{section::scattering} of this work.



\subsection{Laser and material parameters} \label{subsection::Laser_parameter}

We simulate the excitation of a ferromagnetic iron (Fe) layer with a $25\;\mathrm{nm}$ depth with a temporally Gaussian laser pulse with full width half maximum (FWHM) of $25\;\mathrm{fs}$ and $6\;\mathrm{eV}$ photon energy. The simulation works for any spin dependent density of states. Here, we take the data for iron from Ref.~\cite{WEBdataFe}. 

\begin{figure}[!htbp]    
   \includegraphics[scale =0.6, center]{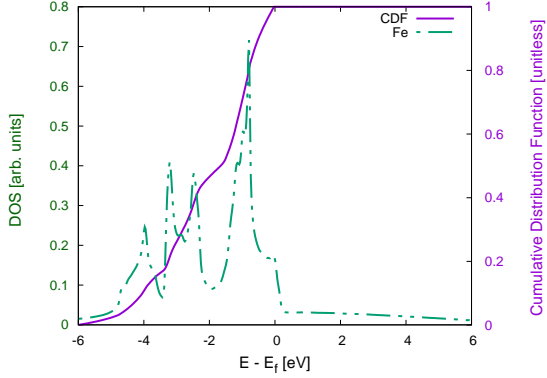}
   \caption{Energy dependent cumulative distribution function (solid line - right axis) behavior when it is weighted according to the spin up density of states (doted line - left axis) of ferromagnetic .}
   \label{fig:DOS_CDF}
\end{figure}

The excitation of an electron from an occupied band is modeled by choosing for eq.~(\ref{eq:sampling}) a probability of excitation according to the material's density of states below $E_\mathrm{F}$. Fig. \ref{fig:DOS_CDF} shows the calculated energy dependent cumulative distribution function (solid line) for the density of states for spin up electrons (dash-dotted line). States above $E_\mathrm{F}$, will be considered as free electron states with a constant effective mass.

\begin{figure}[!htbp]
\centering
   \includegraphics[scale =0.205, center]{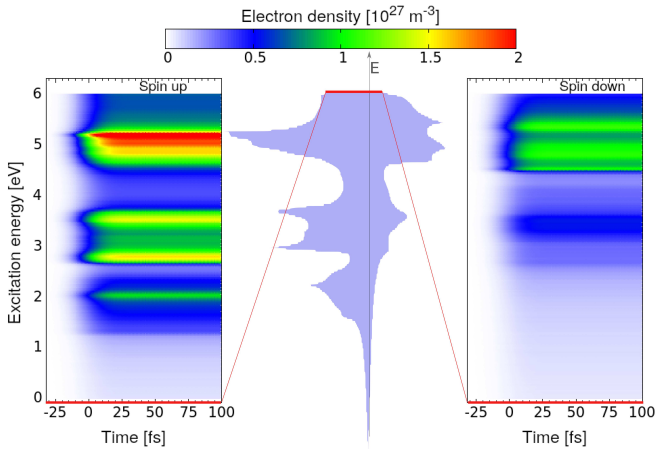}
   \caption{Density of excited electrons vs. time for photoexcitation with a $25\;\mathrm{fs}$ Gaussian laser pulse, but no scattering. Left: Spin up excited electrons. Right: Spin down excited electrons. Middle panel: Sketch of the occupied part the spin-dependent density of states shifted up by the excitation photon energy.}
   \label{fig:DOS}
\end{figure}

As a first test we study only photoexcitation, without any scattering processes for the excited electrons. 
Fig.~\ref{fig:DOS} shows the color-coded electron density of excited spin up (or majority) and spin down (or minority) electrons after photoexcitation in dependence on time and kinetic energy above $E_\mathrm{F}$. 
The laser pulse with $25\;\mathrm{fs}$ duration is centered around $t=0$.
The results show that a larger density of electrons is excited from $3\mathrm{d}^\uparrow$ states in comparison to the $3\mathrm{d}^\downarrow$, i.e., predominantly majority electrons are excited.
The spin-dependent density of states, which is sketched in the background of Fig.~\ref{fig:DOS}, shows that the excited energy-resolved electron density reflects the spin dependent density of states of the occupied bands in ferromagnetic iron.

\section{Different scattering processes}
\label{section::scattering}

\begin{figure}[!htbp]
    \includegraphics[scale =0.4, center]{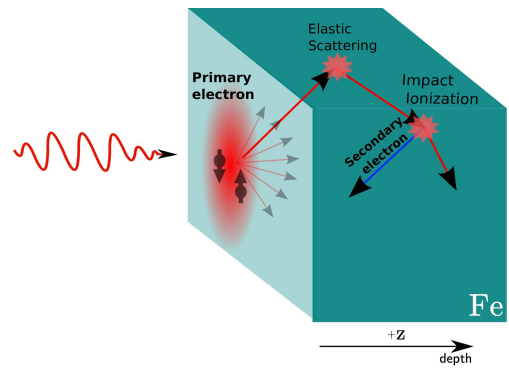}%
    \caption{Schematic figure of hot electron dynamics optically excited. Electrons undergo different interactions in time such as electron-nucleus interaction, where only the direction of the particle is changed (elastic scattering), and electron-electron impact ionization, where secondary electrons are generated (impact ionization).}
    \label{fig:Interact}
\end{figure}

We trace the dynamics of particles after laser excitation by pure jump processes~\cite{FoundationMC, SimulationMC,HUTHMACHER2015242}. In our approach, we consider free electron states above $E_\mathrm{F}$ as essentially free and focus on the influence of high energy electrons in spin transport. Fig. \ref{fig:Interact} shows a schematic representation of how we treat the electron kinetics after ultrafast pulse excitation in a ferromagnet here. During the simulation any electron interaction process is treated by random sampling. For the excitation process already discussed above, the initial direction of the excited electrons will be taken as random direction. 

We next consider the probability that an electron has not suffered any collision between $\tilde t=0$ and $t$ as 

\begin{equation}\label{eq:prob_tau}
    p(t) = \exp\Big[-\int^t_0 \nu\big(E(\tilde t)\big)d\tilde t\Big].
\end{equation}
Here,  $\nu$ is the total scattering rate defined as the sum of all possible scattering transitions. 
It is dependent on the energy $E$ of the considered electrons, which may vary in time. 
In fact, this equation should be solved for each scattering event. 
However, we simplify the calculation as proposed in Ref.~\cite{Rees} by assuming 
a constant scattering rate $\nu_0$. 
Then, the time of free flight $\tau$ can be sampled with the random variable $\mathrm{R} \in [0,1]$ as
\begin{equation}
 \tau = -\nu^{-1}_0\log (\mathrm{R}).
\end{equation}
%
%
However, assuming a constant total scattering rate, independent of energy, 
is a statistical overestimation. 
To compensate, we introduce a further ``collision`` with an energy-dependent probability, which allows
the particle to continue its trajectory unperturbed. 

We randomly sample which collision takes place by using the individual collision frequencies and form the cumulative distribution function in order to solve eq.~(\ref{eq:sampling}). In the following subsections we analyze the interactions that will be considered and the equations or parameters used.

\subsection{Elastic scattering: Electron - nucleus interaction} \label{subsection::elastic}

High-energy electrons are scattered nearly elastically by nuclei, i.e., they change their direction of motion. Here, we parametrize this interaction by two important quantities: The angle of deflection ($\theta$) and the elastic scattering rate ($\tau_{el}^{-1}$). We will follow a procedure which has been applied to different materials, see Refs. \cite{Czyzewski, Gombas, Jablonski}. In order to obtain the angle of deflection $\theta$ for solving eq. (\ref{eq:sampling}) we choose as probability function the differential cross section ($\frac{d\sigma}{d\Omega}$), described here by the Mott cross section \cite{DAPOR1995470}:

\begin{equation} \label{eq:elast}
\frac{d\sigma}{d\Omega} = {|f(\theta)|}^{2} + {|g(\theta)|}^{2},
\end{equation} 
where $\theta$ is the scattering angle, $f(\theta)$ and $g(\theta)$  are the scattering amplitudes which can be obtained from the following expressions:
\begin{eqnarray}
f(\theta) = \frac{1}{2iK} \sum^{\infty}_{l=0} \biggl\{(l+1)&\left[e^{(2i\delta_{l})} - 1\right] + \nonumber \\
&l\left[e^{(2i\delta_{-l-1})} - 1\right] \biggr\} P_{l}(\cos\theta),\\
g(\theta) = \frac{1}{2iK} \sum^{\infty}_{l=1}\biggl\{(l+1)&\biggl[-e^{(2i\delta_{l})} + \nonumber \\
&e^{(2i\delta_{-l-1})}\biggr]\biggr\}P^{1}_{l}(\cos\theta).
\end{eqnarray}
Here, $K^{2}=W^{2}-1$, where $W$ is the total energy of the incident electron in atomic units, $\delta_l$ is the phase shift for the $l$-th partial wave and $P_l$ and $P^{1}_{l}$ are the ordinary and associated Legendre Polynomials, respectively.

We have now a set of equations that will allow us to obtain values of the angle of deflection by using eq.~(\ref{eq:sampling}). 
The next step is to obtain an expression for the elastic scattering rate. 
In principle, it can be deduced from the total scattering cross section, accessible by the integration of eq.~(\ref{eq:elast}), together with the density of collision partners. It will depend on energy, material and temperature \cite{PhysRevB.77.075133, PhysRevB.102.064302,JLA17chen, MT52lee}. 
Since we trace non-interacting electrons in our approach, we have no access to integrated quantities like temperature. 
Therefore, we 
%
focus here on the net effect of the scattering time $\tau_{el}$ between electrons and nuclei by assuming a constant value. 
We will analyze the dynamics for two values: $\tau_{el} = 25\;\mathrm{fs}$ from Ref.~\cite{PhysRevB.71.233104} and $\tau_{el} = 12\;\mathrm{fs}$ from Ref.~\cite{CommCohWilson}. We will not take into account the loss of energy due to recoil for this simulation.


\subsection{Inelastic scattering: Impact Ionization}

When a high-energy electron interacts with electrons below and close to $E_\mathrm{F}$ in the occupied band and generates secondary electrons (SE), we will label this as an inelastic collision. A high-energy primary electron with energy $E$ can lose an amount of energy $\Delta E$ to a second electron, thereby ionizing the latter. 
The SE produced in this process can ionize further secondary electrons in a cascade process. The generation of SE by cascade process is believed to have an important influence on ultrafast spin transport \cite{Eschenlohr}. The process of secondary electrons generation has been studied in different materials, going back to Refs.~\cite{PhysRev.138.A336, wolff}. 

Our numerical treatment of the SE generation process is as follows. In order to select the newly excited electron we follow the same procedure explained in subsection \ref{subsection::Laser_parameter}, where the probability of excitation depends on the material's density of states, with the constraint that two particles cannot be in the same state and thus we avoid selecting two electrons from the same occupied state twice.
We distinguish the spin of the newly excited electrons by selecting them from the spin-resolved density of states and weight the probabilities accordingly.


The energy lost by the primary electron $\Delta E$ is used to ionize a secondary electron from the occupied band. Here we define a binding energy $I$, which is the amount of energy necessary to reach Fermi level from an occupied state. Taking this into account, the final energy of the ionized electron $E_s$ above Fermi level, is then 
$\Delta E = E_s + I$.  
The amount of transferred energy $\Delta E$ is assumed to be 
half of the energy of the incident electron ($E/2$) as it was done in Ref.~\cite{RITCHIE19651689}. 

For the inelastic scattering rate  ($\tau_{ee}^{-1}$) we will use the energy-dependent collision rate~\cite{PhysRevB.87.035139} of an excited electron at temperature $T_{e}$,
\begin{equation}\label{eq:t_ee}
 \frac{1}{\tau_{ee}(E,T_{e})} = \frac{\pi^2 \sqrt{3}}{128} \frac{\omega_{p}}{E_{f}^2} \frac{(\pi k_{B} T_{e})^2 + (E - E_{f})^2}{e^{ \left({-\frac{E - E_{f}}{k_{B}T_{e}}}\right) } + 1}\,,
\end{equation}
where $E_\mathrm{F}$ is the Fermi energy and $\omega_{p}$ is the plasma frequency. 
For the derivation of eq.~(\ref{eq:t_ee}) it is assumed that the band electrons have a Fermi-Dirac distribution. 

For the angle of deflection after an inelastic scattering, we use the classical binary collision model which can be derived from momentum and energy conservation. In collisions between two identical particles with identical masses, the  angles of deflection for the primary ($\theta$) and secondary electron ($\alpha$) are given by
\begin{eqnarray}
 \theta &= \arcsin \left(\sqrt{\frac{\Delta E}{E_s + \Delta E}} \right)\enspace,\\
 \alpha &= \arcsin \left(\sqrt{\frac{E_s}{\Delta E - I}} \sin \theta \right)\enspace.
\end{eqnarray}

Since electrons are not distinguishable, we also include a probability that an electron effectively flips its spin. This is an exchange scattering process, due to the Coulomb interaction with electrons in the occupied part of the spin-split band structure, which has been analyzed in terms of the spin-flip self energy by Hong and Mills~\cite{PhysRevB.62.5589}. While magnons and Stoner excitations can contribute, this scattering processes is called Stoner excitation in Ref.~\cite{mao4823}. We follow that paper and employ the energy-dependent spin-flip probability in our simulation, see Fig.~2 in Ref.~\cite{mao4823}.

\section{Space dependence}\label{spatial}

\begin{figure}[!htbp]
   \includegraphics[scale =0.68, center]{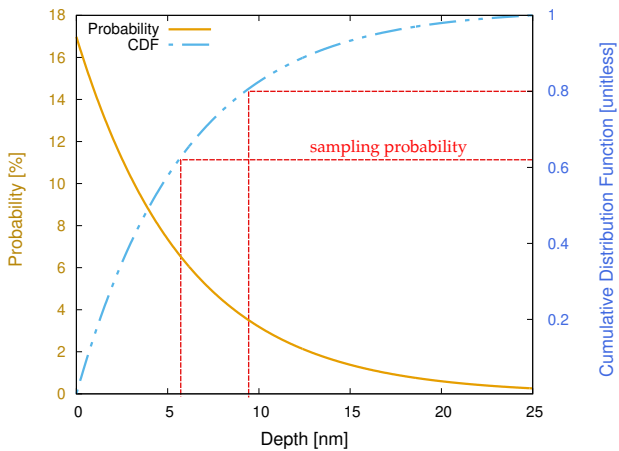}
   \caption{Probability density and cumulative distribution function for electron excitation from penetration depth of the laser pulse according to the Beer-Lambert equation. Red lines represent how the sampling of the initial position of the particles is taken from the values of the Cumulative distribution.}   
   \label{fig:Prob_depth}
\end{figure}

The absorption profile of the optical excitation imprints a spatial dependence on the initial electron distribution. We take this into account by an excitation probability derived from the Beer-Lambert law. We select the initial position of the excited electrons in the material by randomly sampling the position of excitation from the profile shown in Fig.~\ref{fig:Prob_depth}. The initial direction of movement is sampled randomly as well. The kinetic energies are between Fermi energy $E_F$ and 
$E_F + 6\;\mathrm{eV}$, 
according to the excitation of each individual electron. Then, each electron is traced individually in time and we can monitor its displacement throughout the material.


The motion  of electrons can be characterized by the mean square displacement (MSD) \cite{PhysRevE.82.041914}. The MSD is defined as the spatial spread of the distribution, which occurs in the transport direction $z$. As the electrons move, the MSD of their distribution becomes time dependent $ \left<(\Delta z)^2\right> \propto t^{\alpha}$, where $\alpha$ is known as the generalized diffusion exponent. 

The diffusion exponent characterizes the electronic motion under the influence of the excitation and scattering processes. In the ballistic regime, essentially no collisions occur, leading to $\alpha \! =\! 2$. Superdiffusive behavior occurs in the intermediate regime with $2 > \alpha > 1$. On longer timescales, the motion of particles is randomized by multiple scattering processes, leading to a diffusive behavior described by $\alpha = 1$.

\section{Results}

The different scattering mechanisms that may occur after fs-laser excitation in ferromagnets affect the nonequilibrium dynamics. We will focus first on the influence of different scattering rates for elastic scatterings and then discuss the effects of secondary electrons generation. We will analyze the following physical quantities: Displacement of the particles, 
particle velocities, diffusion and spin current.

\subsection{Influence of different elastic scattering times } \label{subsection::Rt_el}

\begin{figure}[!htbp]
   \includegraphics[scale =0.72, center]{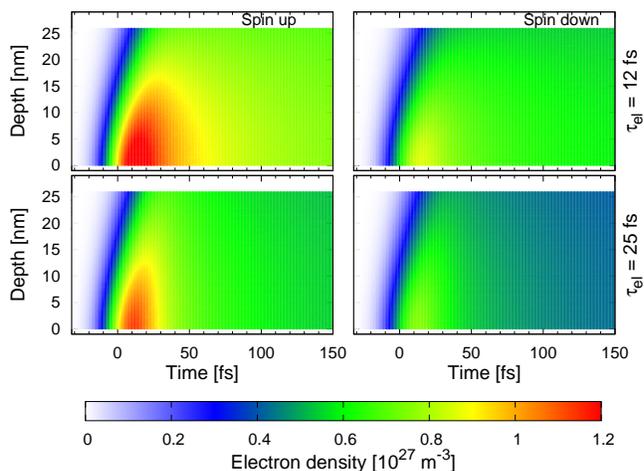}
   \caption{Evolution of particle density through the material for a simulation including all scatterings processes with different scattering times of elastic scatterings: $\tau_{el} = 12\;\mathrm{fs}$ (top figure) and $\tau_{el} = 25\;\mathrm{fs}$ (lower figure) and different spins: Spin up (lhs) and Spin down (rhs).}
   \label{fig:depth_diffSct}
\end{figure}

Figure~\ref{fig:depth_diffSct} shows the time evolution of the particle density in iron with an open boundary at $25\,\mathrm{nm}$ for a simulation where all the scattering processes are included for two different elastic scattering times. The spatio-temporal dynamics of the density of excited electrons is shown in colour code. The surface is at depth zero and the laser is centered at time zero. The upper subplots are for an elastic scattering time of $\tau_{el} = 12\;\mathrm{fs}$, the lower ones for $\tau_{el} = 25\;\mathrm{fs}$. One first observes that for the smaller elastic scattering time (upper subplots) the particles remain close to the surface longer. For the smaller elastic scattering time the excited particle density is observable inside the material for larger times. This is because elastic scattering processes occur more often and change the direction of the particles, contributing to the spreading in the material. 
The red areas in the spin-up channel indicate a higher density of excited spin-up electrons, which is due to the band structure features discussed in the previous section \ref{section::excitation} on the photoexcitation process. 
After about 75 fs 
the signature of the spatial laser penetration profile has
been washed out by scattering processes and transport. The transport characteristics of the dynamics shown here will be analyzed in more detail using the mean square displacement (MSD) in the following subsection.

\begin{figure}[!htbp]
\centering
   \includegraphics[scale =0.68, center]{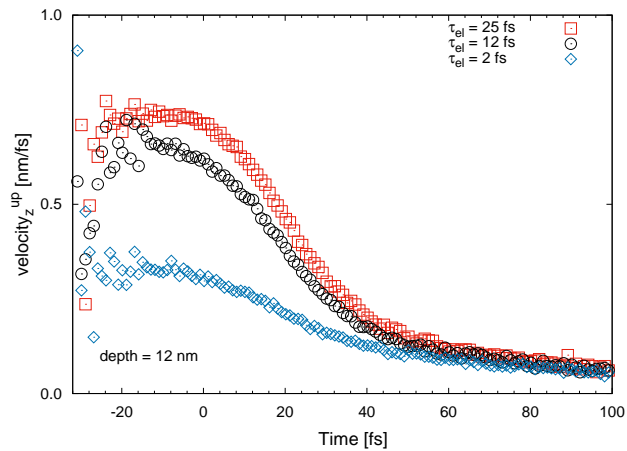}
   
   \caption{ Mean velocity of particles with spin up in z-direction at $12\;\mathrm{nm}$ for different elastic scattering times ($\tau_{el}$). The graph for spin down electrons was omitted since it showed a similar tendency.}
   
   \label{fig:Vel}
\end{figure}

 Figure~\ref{fig:Vel} shows the mean velocity in $z$ direction at a depth of $12\,\mathrm{nm}$. Both kinds of spin show only statistical differences, here we show the results for spin up electrons. The mean velocity is calculated as the average velocity in $+z$ direction of all excited particles at the given depth. Apart from the scattering times discussed in subsection \ref{subsection::elastic} we also show results for a third, shorter one ($\tau_{el} = 2\;\mathrm{fs}$) as a way to observe how the system is influenced by lower values of $\tau_{el}$. During the first femtoseconds one can observe higher average velocities with direction into the depth of the material ($\mathrm{velocity}_\mathrm{z}^{\mathrm{up}} > 0$), but decreasing in magnitude for lower elastic scattering times. 
 In all three cases, the different scattering times keep the velocity of particles on average pointing into the material. 
 After $60\;\mathrm{fs}$ the average velocity for all three cases approaches the same value and continues decreasing at the same rate but without reaching zero, this means that a large number of particles travel in $+z$ direction. During the first femtoseconds a smaller magnitude of average velocities towards the depth can be observed for lower elastic scattering time. This is due to a larger number of scatterings that occur influencing the dynamics of electrons traveling through the solid.

\subsection{Influence of secondary-electron generation}

\begin{figure}[!htbp]
   \includegraphics[scale =0.7, center]{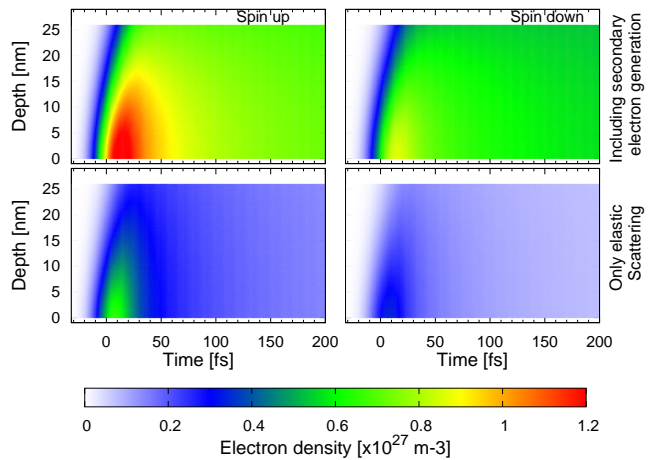}
   
   \caption{Evolution of particle density of spin up (lhs) and spin down (rhs)  particles in the material. Lower figure: Particles travel through the material and they change their direction of flight only (Elastic scattering). Top figure: Particles travel experiencing two scatterings, elastic scattering and impact ionization which generates secondary electrons. Simulation for $\tau_{el}=12\;\mathrm{fs}$.}
   
   \label{fig:depthE}
\end{figure}

For the analysis of the influence of secondary electrons (SE) we compare calculations with and without SE. In Fig.~\ref{fig:depthE} the displacement of particle density in the material for spin up (left hand side) and spin down (right hand side) is presented. The lower subplots shows calculations without including secondary electrons, labeled "elastic scattering", whereas the upper subplots shows simulations including secondary electrons generation, labeled "inelastic scattering" (it is the same as the upper panel of Fig.~\ref{fig:depth_diffSct}, repeated here for convenience). During the first femtoseconds, whether secondary electrons are generated or not, one observes a larger concentration of particles near the surface. Later in time particles spread fast from the surface into the material because more scatterings take place. This indicates that the generation of SEs increases the spread of the particles into the material. The increase in displacement throughout the material can be examined better with the analysis of the motion regimes using the mean square displacement (MSD).

\begin{figure}[!htbp]
   \includegraphics[scale =0.7, center]{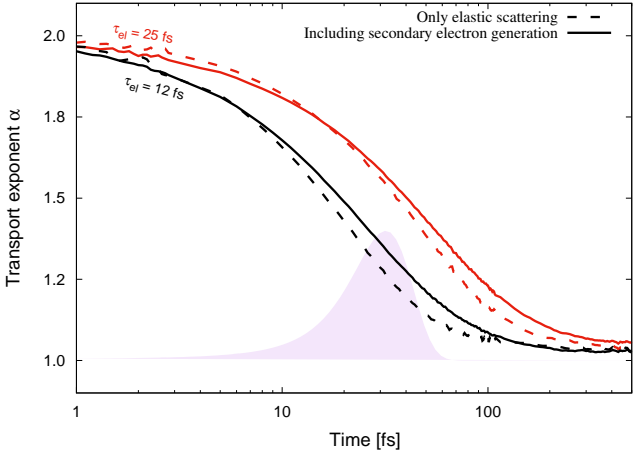}
   \caption{Analysis of the transport exponent $\alpha$ when using different elastic scattering times and the influence of secondary electrons for spin up electrons. When the system has a lower elastic scattering time $\tau_{el}$ it relaxes faster into the diffusive regime whereas secondary electrons make this transition longer.}
   \label{fig:MSD_diffSct}
\end{figure}

Figure~\ref{fig:MSD_diffSct} shows a comparison in the evolution of the transport exponent $\alpha$ for two different elastic scattering times $\tau_{el}$ with and without the inclusion of secondary electron generation. Only the analysis for spin up electrons is shown since the spin down electrons present on average a similar behavior with only slightly different magnitudes. The data from Figs.~\ref{fig:depth_diffSct} and \ref{fig:depthE} are analyzed now
as described in section \ref{spatial} using the MSD with the transport exponent $\alpha$. One can observe for $\tau_{el}=12\;\mathrm{fs}$ and $\tau_{el}=25\;\mathrm{fs}$ distinctively all three motion regimes. Starting from ballistic, going through superdiffusive and finally becoming diffusive. Since the particles can in principle be initially excited with an arbitrary initial direction pointing into the material, in Fig. \ref{fig:MSD_diffSct} during the first femtoseconds the motion is not entirely ballistic.

The 
approach of
the diffusive regime occurs at different times, it is faster for smaller scattering time. When the secondary electrons come into play, transition from superdiffusive into diffusive regime is delayed. The generation of secondary electrons 
effectively increases the duration of electron excitation, influencing the system and keeping it in the superdiffusive regime ($\alpha > 1$) for a larger time in comparison with the other calculations. These results are in agreement  with those in Refs.~\cite{PhysRevB.98.224416, PhysRevB.86.024404}.  

\begin{figure}[!htbp]
   \includegraphics[scale =0.7, center]{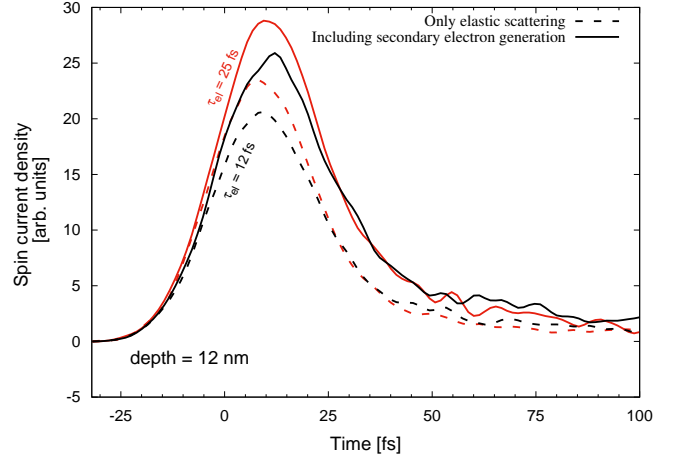}
   \caption{Spin current density at a depth of $12\;\mathrm{nm}$ for different $\tau_{el}$ for a simulation where it is compared two modeling assumptions: Including generation of secondary electrons (solid line) and not including them (dotted line). }
   \label{fig:current}
\end{figure}

In the study of spin transport, spin current density is one of the main features to be analyzed. The spin current density $j_s$ is defined as
\begin{equation}
    j_s(z,t)\propto q[ \langle \eta^{\uparrow} v_\uparrow \rangle - \langle \eta^{\downarrow} v_\downarrow \rangle ] \,,
\end{equation}
where $q$ is the charge of the electron, $\eta^{\uparrow}$ ($\eta^{\downarrow}$) and $v_\uparrow$ ($v_\downarrow$) are the particle density and the velocity for spin up (spin down), respectively. With this definition the spin current is positive if effectively more spin up electrons move into positive $z$ direction. The spin current density at a fixed depth of 12 nm with (solid line) and without (dashed line) secondary electron generation is shown in Fig.~\ref{fig:current}. One can observe that the spin current density changes quantitatively due to the continuous generation of secondary electrons, which feeds  excited electrons into the dynamics. Fig.~\ref{fig:current} shows also a change in  the time of maximum intensity in the spin current density when secondary electrons are generated. As a result, the propagation time is extended. In the bulk of the ferromagnet, the spin current does not change sign during the whole simultation. We note that this is different from spin-polarized transport in normal metals where the excitation conditions together with the transport characteristics can lead to a bipolar spin-current signal~\cite{PhysRevB.98.224416}.

\section{Summary}

In conclusion, we developed a kinetic Monte Carlo method to study the influence of different electron-nucleus collision rates and generation of secondary electrons in the ultrafast nonequilibrium spin and charge transport in Iron. This method simulates kinetics of individual particles based on random sampling, making it a powerful tool for tracing electrons throughout the material. In this simulation we used the probability of excitation according to the material's density of state to excite and electron from an occupied band. The displacement of the particles in the material were plotted, along with the mean velocity ($\mathrm{velocity}_\mathrm{z}^{\mathrm{up}}$) to analyze the dynamics of excited electrons for different elastic scattering times at certain depth. We found that lower scattering times increase the average velocities pointing into the depth. Regarding the influence of the secondary electron generation, we have focused on their impact in different regimes of motion and spin current density. Generation of secondary electrons effectively delay the excitation of free electrons. Therefore, 
they influence the system by delaying the transition from one motion regime to another. They also affect the intensity of the spin current density and change the time of its maximum peak.
We conclude that the spin dynamics is determined by the elastic scattering time as well as by the generation of secondary electrons. 
Both studied variations influence different transport parameters which can affect experimental observations. 


\section*{References}

\bibliographystyle{iopart-num}
\bibliography{biblio}

\end{document}